\documentclass[12pt]{article}
\pdfoutput=1
\usepackage{epsf,amsfonts,amssymb,epsfig,amsmath,mathtools,graphics,slashed}
\usepackage{hyperref,hep,graphicx,subfig}

\makeatletter

\newcommand{\Rmnum}[1]{\expandafter\@slowromancap\romannumeral #1@}
\makeatother

\newcommand{\nn}{\notag \\}

\begin{document}

\makeatletter
\renewcommand{\theequation}{\thesection.\arabic{equation}}
\@addtoreset{equation}{section}
\makeatother

\baselineskip 18pt

\begin{titlepage}

\vfill

\begin{flushright}
Imperial/TP/2013/JG/04\\
\end{flushright}

\vfill

\begin{center}
   \baselineskip=16pt
   {\Large\bf Holographic Q-lattices}
  \vskip 1.5cm
  \vskip 1.5cm
      Aristomenis Donos$^1$ and Jerome P. Gauntlett$^2$\\
   \vskip .6cm
    \vskip .6cm
      \begin{small}
      \textit{$^1$DAMTP, 
       University of Cambridge\\ Cambridge, CB3 0WA, U.K.}
        \end{small}\\
      \vskip .6cm
      \begin{small}
      \textit{$^2$Blackett Laboratory, 
        Imperial College\\ London, SW7 2AZ, U.K.}
        \end{small}\\*[.6cm]

\end{center}

\vfill

\begin{center}
\textbf{Abstract}
\end{center}

\begin{quote}
We introduce a new framework for constructing black hole solutions that are holographically dual 
to strongly coupled field theories with explicitly broken translation invariance. 
Using a classical gravitational theory with a continuous global 
symmetry leads to constructions that involve solving ODEs instead of PDEs. 
We study in detail $D=4$ Einstein-Maxwell theory coupled to
a complex scalar field with a simple mass term. 
We construct black holes dual to metallic phases which exhibit a Drude-type peak in the
optical conductivity, but there is no evidence of an intermediate scaling that 
has been reported in other holographic lattice constructions.
We also construct black holes dual to insulating phases which exhibit a suppression of spectral weight at low frequencies.
We show that the model also admits a novel $AdS_3\times\mathbb{R}$ solution.
\end{quote}

\vfill

\end{titlepage}
\setcounter{equation}{0}


\section{Introduction}
It is a remarkable fact that many phenomena observed in condensed matter systems are now known to have gravitational 
analogues via the AdS/CFT correspondence. One area of focus, where there has been significant recent progress, concerns the
holographic description of physics associated with a ``lattice". More specifically, there are are now several different constructions of black hole solutions
that are holographically dual
to strongly coupled systems which explicitly break translation invariance using a spatially periodic deformation
\cite{Horowitz:2012ky,Horowitz:2012gs,Horowitz:2013jaa,Donos:2012js,Ling:2013nxa,Chesler:2013qla}.

One motivation for constructing such black holes arises in the context of studying the optical conductivity of strongly coupled systems
at finite charge density. 
In the absence of a lattice the translation invariance of the system implies
that there is a delta function peak at zero frequency, implying that the system is an ideal conductor.
To extract more realistic metallic behaviour one can investigate the impact of a lattice. The first construction of
electrically charged black holes describing holographic lattices was made in $D=4$ Einstein-Maxwell theory coupled to
a real scalar field \cite{Horowitz:2012ky}. For the specific black holes that were constructed, it was shown that the system is in a metallic phase 
with the delta function peak smeared out into a Drude-type peak\footnote{Drude-type physics has also been discussed in a holographic context in, for example, \cite{Karch:2007pd,Hartnoll:2007ih,Hartnoll:2007ip,Hartnoll:2009ns,Faulkner:2010zz,Hartnoll:2012rj,Liu:2012tr,Vegh:2013sk,Mahajan:2013cja,Davison:2013jba,Faulkner:2013bna}.}.
This observed low frequency behaviour is consistent with the 
general analysis of conductivities that was made earlier in \cite{Hartnoll:2012rj} (see also \cite{Mahajan:2013cja}).

Moving away from the low-frequency regime, with the scale set by the chemical potential,
a particularly striking conclusion of \cite{Horowitz:2012ky} was that the optical conductivity appears to exhibit 
a power-law behaviour at intermediate frequencies. More precisely the optical conductivity was seen to have
the form  
\begin{align}\label{scal}
|\sigma(\omega)|=B\omega^{-2/3}+C\,,
\end{align}
where $B,C$ are frequency independent constants, 
and furthermore, the same behaviour was also seen for other lattices and other spacetime dimensions in \cite{Horowitz:2012gs,Horowitz:2013jaa,Ling:2013nxa}. Since an intermediate scaling of the optical conductivity for the high $T_c$
cuprates is seen with the same scaling exponent $-2/3$, albeit with $C=0$
and a frequency independent phase (e.g. \cite{2003Natur.425..271M,2006AnPhy.321.1716V}),
it is important to analyse this result in more detail. In fact for the holographic lattice that we construct in
this paper we will not see such scaling behaviour. 
We will discuss the connection between our results and 
\cite{Horowitz:2012ky,Horowitz:2012gs,Horowitz:2013jaa,Ling:2013nxa} at the end of the paper.

A more recent motivation for studying holographic lattices is that it provides a framework for 
investigating metal-insulator transitions within a holographic context \cite{Donos:2012js}. 
This is particularly interesting because there are many perplexing
systems, such as the cuprates, where such transitions are observed and holographic techniques may provide important new insights.
The strategy of \cite{Donos:2012js}
is to construct black holes dual to holographic lattices that flow in the IR to metallic ground states and then to vary the strength and/or the 
periodicity of the lattice aiming to induce a  transition to a new insulating phase. In \cite{Donos:2012js} this was achieved using 
$D=5$ electrically charged black holes dual to helical lattices. Furthermore, new zero temperature insulating ground states that
break translation invariance were also found in \cite{Donos:2012js}.

An important technical issue that arises in constructing black holes dual to lattices is that, in general, they require solving partial differential equations.
For example, the holographic lattices that were constructed in \cite{Horowitz:2012gs,Horowitz:2013jaa,Horowitz:2012ky,Ling:2013nxa} 
break translation invariance in one of the spatial dimensions and lead to a problem in PDEs in two variables; the one spatial direction as
well as a radial direction. For the general setup where the translation invariance is broken in all of the spatial directions, 
time independent black holes in $D$ spacetime dimensions will typically depend on $D-2$ spatial variables as well as a radial variable, leading to PDEs in $D-1$ variables. For $D=4,5$ solving such PDEs numerically is an involved exercise. 
An interesting exception is the construction of the $D=5$ black holes dual to helical
lattices \cite{Donos:2012js}, where a Bianchi VII$_0$ symmetry was utilised to construct black holes by solving ODEs only. 

In this paper we introduce a new framework for constructing holographic lattices that also involves just solving ODEs. 
The key idea is to break the translation invariance by
exploiting a continuous global symmetry
of the bulk classical gravitational theory. A simple theory that can be used to illustrate the idea, 
which is also the theory we will
focus on in the paper, consists of
Einstein-Maxwell theory coupled to a complex scalar field, $\phi$.
The field $\phi$ is neutral with respect to the Maxwell field, and the model is taken to have a global $U(1)$ symmetry 
in addition to the $U(1)$ gauge-symmetry associated with the Maxwell field. 
For example, the Lagrangian density involving $\phi$ can take the form
\begin{align}
{\cal L}(\phi)=\sqrt{-g}\left[-|\partial\phi|^2-V(|\phi|)\right]\,,
\end{align}
leading to the following contribution to the bulk stress-tensor
\begin{align}\label{stten}
T_{\mu\nu}(\phi)=\partial_{(\mu}\phi\partial_{\nu)}\phi^*-\frac{1}{2}g_{\mu\nu}\left[|\partial\phi|^2+V(|\phi|)\right]\,.
\end{align}
The breaking of the translation invariance in, say, the $x_1$ direction can be achieved using the ansatz $\phi=e^{ikx_1}\varphi(r)$ and it is clear
from the form of the stress tensor given in \eqref{stten} that this can be combined with an ansatz for the metric and Maxwell fields that is dependent on the radial
variable only\footnote{In the process of writing up this work, this possibility was also pointed out in a footnote in \cite{Blake:2013owa}.}. This construction shares some similarities with the construction of Q-balls \cite{Coleman:1985ki}, which exploits a 
global symmetry and a time dependent phase to construct spherically
symmetric solitons, 
and so we call them holographic Q-lattices. 

It is worth noting that this particular Q-lattice, involving a single complex scalar field, can be viewed as arising from two real scalar 
fields, with the same mass, each with a periodic spatial dependence in the same direction that is shifted by an amount $\pi/2k$. In this sense
it can be viewed as a simple generalisation of the lattice studied in \cite{Horowitz:2012gs}.
More generally, this lattice construction can easily be extended to study the breaking of translation invariance in additional spatial directions by considering a model
with a larger global symmetry. For example, one can use a model with additional complex scalar fields and with additional global $U(1)$ symmetries. 
One can also have larger global symmetry groups and/or use higher rank tensor fields instead of scalars.
Such lattices will be studied in detail elsewhere.

The plan of the rest of the paper, including some of the key results, are as follows.
In section \ref{bhs} we study $D=4$ Einstein-Maxwell theory coupled to a complex scalar field with a simple mass term.
We construct Q-lattice black holes that describe metallic phases which at zero temperature approach $AdS_2\times\mathbb{R}^2$ 
in the far IR. We numerically calculate the low temperature behaviour of the DC resistivity and extract the scaling  
behaviour that is predicted from \cite{Hartnoll:2012rj} using the memory matrix formalism. This comprises the first\footnote{We will comment on the results of \cite{Horowitz:2012ky} in section \ref{fincom}.} numerical confirmation of 
\cite{Hartnoll:2012rj} for fully back reacted black holes and complements the recent analytic results of \cite{Blake:2013owa}
in the context of perturbative lattices.
We also construct black holes that describe insulating phases, realising the first holographic metal-insulator 
transition for $d=3$ field theories. At low temperatures there is a transfer of spectral weight in the insulating phase and the real part of the optical conductivity develops
a mid frequency hump. Some details of the conductivity calculation is presented in section \ref{condsec}, which includes some new technical material. 
Interestingly, the model that we analyse also admits an $AdS_3\times\mathbb{R}$ solution which we discuss in an appendix.
We conclude with some final comments in
section \ref{fincom}, including a discussion of the absence of intermediary scaling in the optical conductivity.

\section{Black hole solutions}\label{bhs}
We shall consider $D=4$ Einstein-Maxwell theory coupled to a complex field $\phi$ with action given
by
\begin{align}\label{act}
S=\int d^4 x\sqrt{-g}\left[R+6-\frac{1}{4}F^2-|\partial\phi|^2 -m^2|\phi|^2\right]\,,
\end{align}
where $F=d A$. We have set $16\pi G=1$ and also fixed the scale of the cosmological constant for convenience.
The equations of motion can be written
\begin{align}\label{eqsmot}
R_{\mu\nu}&=g_{\mu\nu}(-3+\frac{m^2}{2}|\phi|^2)+\partial_{(\mu}\phi\partial_{\nu)}\phi^*+\tfrac{1}{2}\left( F^2_{\mu\nu}-\tfrac{1}{4}g_{\mu\nu}F^2\right)\,,\nn
&\nabla_\mu F^{\mu\nu}=0,\qquad
(\nabla^2-m^2)\phi=0\,,
\end{align}
and admit an $AdS_4$ vacuum solution, with unit radius, which is dual to a $d=3$ CFT. The CFT has two global abelian symmetries.
The first arises from the gauge symmetry in the bulk and there is a corresponding conserved current which is dual to the bulk-gauge field $A$. 
The second arises from the global symmetry in the bulk, associated with multiplying $\phi$ by a constant phase, and there is not a
corresponding conserved current\footnote{A discussion of such global symmetries arising in a different holographic context appears in \cite{Amado:2013xya}.}
in the CFT. 
The CFT also has a complex scalar operator 
with scaling dimension $\Delta= 3/2\pm (9/4+m^2)^{1/2}$ dual to the scalar field $\phi$. We want this to be a relevant operator
in a unitary CFT and hence we take $-9/4\le m^2<0$.

The CFT at finite temperature $T$ and chemical potential $\mu$ can be holographically  
described by the
standard electrically charged AdS-RN black solution given by
\begin{align}
ds^2&=-Udt^2-U^{-1}dr^2+r^2\left(dx_1^2+dx_2^2\right)\,,\nn
A&=\mu(1-\frac{r_+}{r})dt\,,
\end{align}
with $\phi=0$ and $U=r^2-(r_+^2+\frac{\mu^2}{4})\frac{r_+}{r}+\frac{\mu^2r_+^2}{4r^2}$.
The temperature is given by $T=(12r_+^2-\mu^2)/16\pi r_+$ 
and at $T= 0$ it approaches 
the following $AdS_2\times\mathbb{R}^2$ solution as $r\to r_+$:
\begin{align}\label{ads2}
ds^2&=\frac{1}{6}ds^2(AdS_2)+dx_1^2+dx_2^2\,,\nn
F&=\frac{1}{\sqrt{3}}Vol(AdS_2)\,,
\end{align}
where $ds^2(AdS_d)$ denotes the standard unit radius metric on $AdS_d$.

For the mass window $-9/4\le m^2<-3/2$ the scalar field $\phi$ violates the
$AdS_2$ BF bound and hence the AdS-RN black hole solution will become unstable at some
temperature, leading to a different $T=0$ ground state. In order to exclude this possibility,  
for most of the paper we will consider 
\begin{align}
m^2=-\frac{3}{2}\qquad \leftrightarrow\qquad  \Delta=\frac{3+\sqrt 3}{2}\,.
\end{align}
At the end of the paper we will comment on the case $m^2=-2$ and $\Delta=2$.

\subsection{Black hole ansatz for the holographic $Q$-lattice}
We are interested in describing the $d=3$ CFT with chemical potential $\mu$ and an explicit breaking of 
translation invariance in one of the spatial directions, which we take to be $x_1$.
The ansatz we shall consider is given by
\begin{align}\label{ansatzbh}
ds^2&=-Udt^2+U^{-1}dr^2+e^{2V_1}dx_1^2+e^{2V_2}dx_2^2\,,\nn
A&=adt\,,\nn
\phi&=e^{ikx_1}\varphi\,,
\end{align}
where $U, V_1, V_2, a$ and $\varphi$ are functions of the radial co-ordinate only and $k$ is a constant.
Substituting this ansatz into \eqref{eqsmot} we find that the equations of motion can be equivalently recast as 
four second order ODEs for $V_1, V_2, a, \varphi$ and one first order ODE for $U$. It is useful to note that this ansatz is invariant
under the scaling $t\to ct, x_i\to c x_i,r\to c^{-1}r$ and $U\to c^{-2}U, e^{V_i}\to c^{-1}e^{V_i}, a\to c^{-1}a, k\to c^{-1}k$.

We will impose the following boundary conditions on the ODEs.
We demand that we have a regular solution at the black hole event horizon at $r=r_+$, which leads to an expansion depending on
six independent constants $r_+, V_{1+}, V_{2+},V_{22},a_+$ and $\varphi_+$. Specifically as $r\to r_+$ we have
\begin{align}
U&=4\pi T(r-r_+)+\dots,\nn
V_1&=V_{1+}+\left(1-\frac{4e^{-2V_{1+}}\varphi_+^2 k^2}{12-a_+^2-2\varphi_+^2 m^2}\right)V_{22}(r-r_+)\dots, \nn
V_2&=V_{2+}+V_{22}(r-r_+)\dots, \nn
 a&=a_+(r-r_+)+\left(-1+\frac{2e^{-2V_{1+}}\varphi_+^2 k^2}{12-a_+^2-2\varphi_+^2 m^2}\right)a_+V_{22}(r-r_+)^2\dots,\nn
\varphi&=\varphi_++\frac{4(m^2+e^{-2V_{1+}}k^2)}{12-a_+^2-2\varphi_+^2 m^2}\varphi_+V_{22}(r-r_+)\dots,
\end{align}
where $T$ is the temperature of the black hole given by
\begin{align}
T=(4\pi)^{-1}\frac{12-a_+^2-2\varphi_+^2 m^2}{4V_{22}}\,.
\end{align}

At the UV boundary, $r\to\infty$, we demand that we approach $AdS_4$ with deformations corresponding to chemical potential
$\mu$ and lattice deformation parameter $\lambda$. We find that, schematically, we can develop the expansion
\begin{align}\label{uvexp}
U&=r^2+\dots-\frac{M}{r}+\dots,\nn
 V_1&=\log r+\dots+\frac{V_v}{r^3}+\dots,\nn
 V_2&=\log r+\dots-\frac{V_v}{r^3}+\dots,\nn
a&=\mu+\frac{q}{r}\dots,\nn
\varphi&=\frac{\lambda}{r^{3-\Delta}}+\dots +\frac{\varphi_c}{r^{\Delta} }+\dots\,.
\end{align}
This gives a UV expansion that depends on seven parameters $M, V_v, \mu,q,\lambda,\varphi_c$ and $k$.

Notice that for fixed $m^2$, the holographic Q-lattice is specified by three dimensionless quantities fixing the deformations in the UV: 
$T/\mu$, $\lambda/\mu^{3-\Delta}$ and $k/\mu$. We thus expect a three-parameter family of black holes. We have four second order ODEs and one first order ODE, and so a solution is specified by nine parameters. We have six parameters for the near horizon expansion plus another seven for the UV expansion. After subtracting one for the scaling symmetry that the system of ODEs possesses, we deduce that there is indeed, generically, a three-parameter family of black hole solutions.
We also note that the scaling symmetry can be used to set $\mu=1$ if one wishes.

We will choose specific values in the two-dimensional space parameterised by $\lambda/\mu^{3-\Delta}$ and $k/\mu$, and
then examine the behaviour as $T/\mu$ is lowered. In particular, we will see that there is a transition
from metallic to insulating behaviour as we move in this two-dimensional space.

\subsection{Black holes dual to the metallic phase}
The CFT deformed by the Q-lattice will be in a metallic phase if the zero temperature limit of the black hole solutions
interpolate between the lattice deformed $AdS_4$ in the UV and the stable $AdS_2\times \mathbb{R}^2$ solution in the IR.
Indeed this will happen when the lattice deformation in the UV becomes an irrelevant deformation of the
$AdS_2\times \mathbb{R}^2$ solution in the IR, and then the general arguments of \cite{Hartnoll:2012rj}, based on the memory matrix formalism, 
show
that the $T=0$ ground state must be metallic. In particular, at low temperatures, $T<<\mu$, the DC 
resistivity is expected to scale as\footnote{Note that a different, non-standard, definition of $\Delta(k)$ is used in
\cite{Hartnoll:2012rj,Donos:2012js,Blake:2013owa} for this expression.}
\begin{align}\label{dares}
\rho\sim \left(\frac{T}{\mu}\right)^{2\Delta(k)-2}\,,
\end{align}
where $\Delta(k)$ is the smallest scaling dimension of the $k$-dependent irrelevant operators in the locally quantum critical theory
arising in the IR. 
In addition to $k$, $\Delta(k)$ depends on other UV data, as we discuss below.
Furthermore, there should be a Drude peak in the optical conductivity at small temperatures, which at $T=0$ becomes
a delta-function at zero frequency. 

To examine when this situation can arise we now analyse perturbations about the $AdS_2\times\mathbb{R}^2$
solution. Within our ansatz we consider
\begin{align}
U&=6r^2(1+u_1 r^\delta),\quad V_1=v_{10}(1+v_{11}r^\delta),\quad V_2=v_{20}(1+v_{21}r^\delta),\nn
 a&=2\sqrt{3}r(1+a_1 r^\delta),\quad\phi=e^{ikx_1}\varphi_1 r^\delta\,.
\end{align}
The corresponding perturbations are associated with operators with scaling dimension $\Delta=-\delta$ or $\Delta=\delta+1$
in the locally quantum critical IR theory captured by the $AdS_2\times\mathbb{R}^2$ solution. 
We find after substituting into equations of motion
the exponents come in four pairs, satisfying $\delta_++\delta_-=-1$, with $\delta_+=0,0,1$ and a mode just
involving the scalar field with $\delta_+=\delta_\varphi$, where
\begin{align}\label{deltexp}
\delta_\varphi
=-\frac{1}{2}+\frac{1}{2\sqrt 3}\sqrt{3+2m^2+2 e^{-2v_{10}}k^2}\,.
\end{align}
There is also another additional single mode with $\delta_+=-1$ (corresponding to $r_+$ in \eqref{adsdwexp} below).
What is most significant here is that the scalar field perturbation will be an irrelevant deformation 
in the IR (i.e. $\delta_\varphi>0$), provided that the lattice deformation in the IR satisfies
\begin{align}\label{kcond}
(e^{-v_{10}}k)^2>-m^2\,.
\end{align}
In this case the dimension of the irrelevant operator in the locally quantum critical theory is given by $\Delta(k)=1+\delta_\varphi$ and we have
\begin{align}\label{dimk}
\Delta(k)=\frac{1}{2}+\frac{1}{2\sqrt 3}\sqrt{3+2m^2+2 e^{-2v_{10}}k^2}\,.
\end{align}

When \eqref{kcond} is satisfied we can use the two marginal modes with $\delta_+=0$ and the two irrelevant modes to construct domain
walls interpolating between the lattice deformed $AdS_4$ in the UV and the $AdS_2\times\mathbb{R}^2$ solution in the IR.
Specifically, we can develop the following IR expansion
\begin{align}\label{adsdwexp}
U&=6(r-r_+)^2(1-\frac{4}{3v_{10}}V_{+}(r-r_+)+\dots)\,,\nn
V_1&=v_{10}(1+V_+(r-r_+)+\dots)\,,\nn
V_2&=v_{20}(1+\frac{v_{10}}{v_{20}}V_+(r-r_+)+\dots)\,,\nn
a&=\sqrt{12}(r-r_+)(1-v_{10}V_+\dots)\,,\nn
\varphi&=\varphi_+(r-r_+)^{\delta_\varphi}+\dots\,.
\end{align}
We have five IR parameters, $r_+, v_{10},v_{20},V_+,\varphi_+$ and hence when combined with
the UV expansion \eqref{uvexp} and taking into the scaling symmetry, we expect, generically, a two parameter family
of solutions which can be labelled by $\lambda/\mu^{3-\Delta}$ and $k/\mu$.

For the values of $\lambda/\mu^{3-\Delta}$, $k/\mu$ where these domain walls exist, we expect that 
they will arise as the zero temperature limit of lattice deformed black holes which will have, for very small $T/\mu$,
DC resistivity scaling as in \eqref{dares} and a Drude peak in the optical conductivity for small $\omega/\mu$, of the form
\begin{align}\label{drude}
\sigma\sim \frac{K\tau}{1-i\omega \tau}\,,
\end{align}
for constant $K,\tau$.
It should be stressed that the value of $\Delta(k)$ appearing in the DC resistivity depends on 
the value of $v_{10}$ which is fixed by the details of domain wall solution, including all UV data. 
In effect the value of $v_{10}$ is 
renormalising the lattice momentum from $k$ in the UV to $e^{-v_{10}}k$ in the IR.

\begin{figure}
\centering
\subfloat[]{\includegraphics[width=7.3cm]{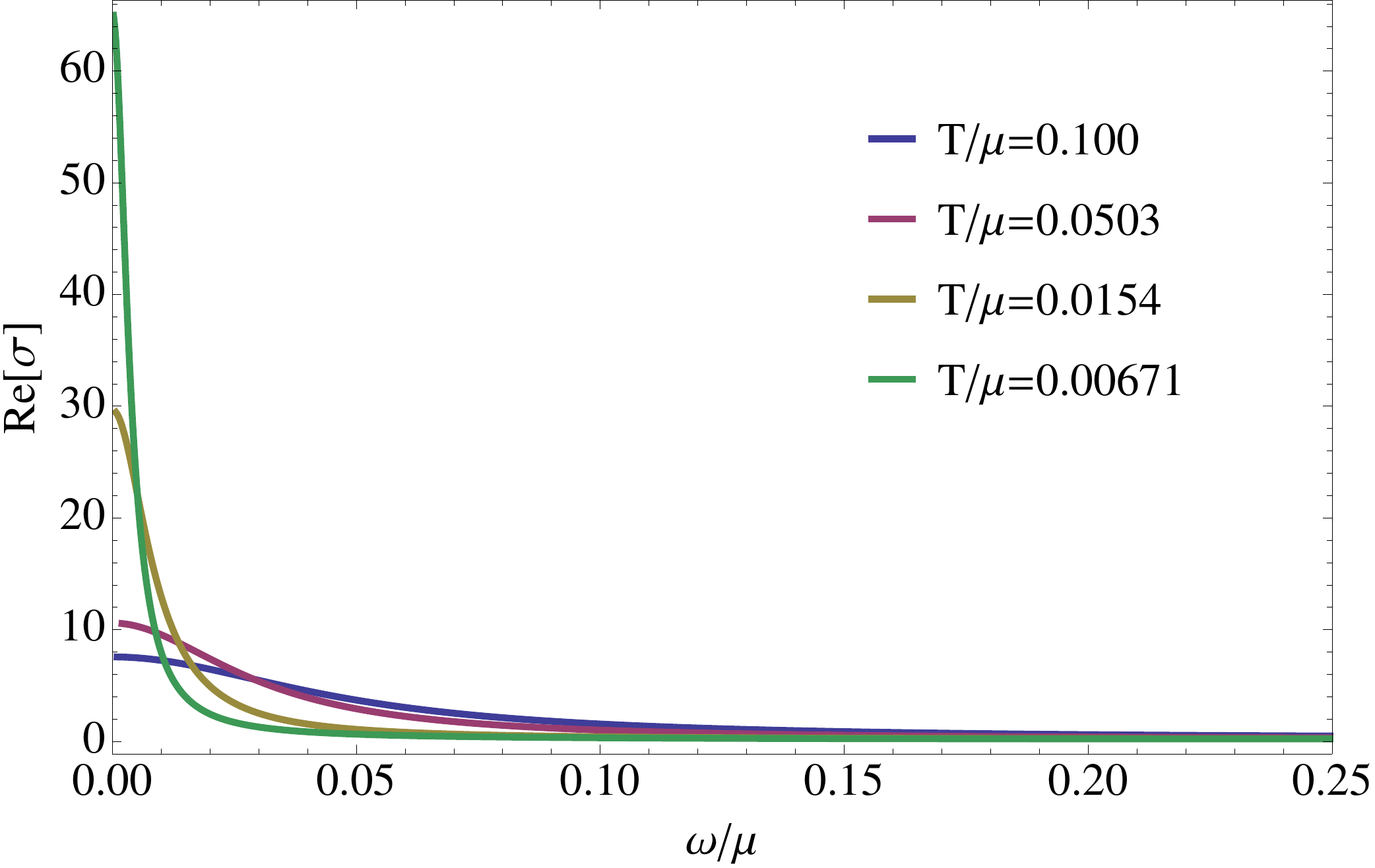}}\hskip 1 em
\subfloat[]{\includegraphics[width=7.3cm]{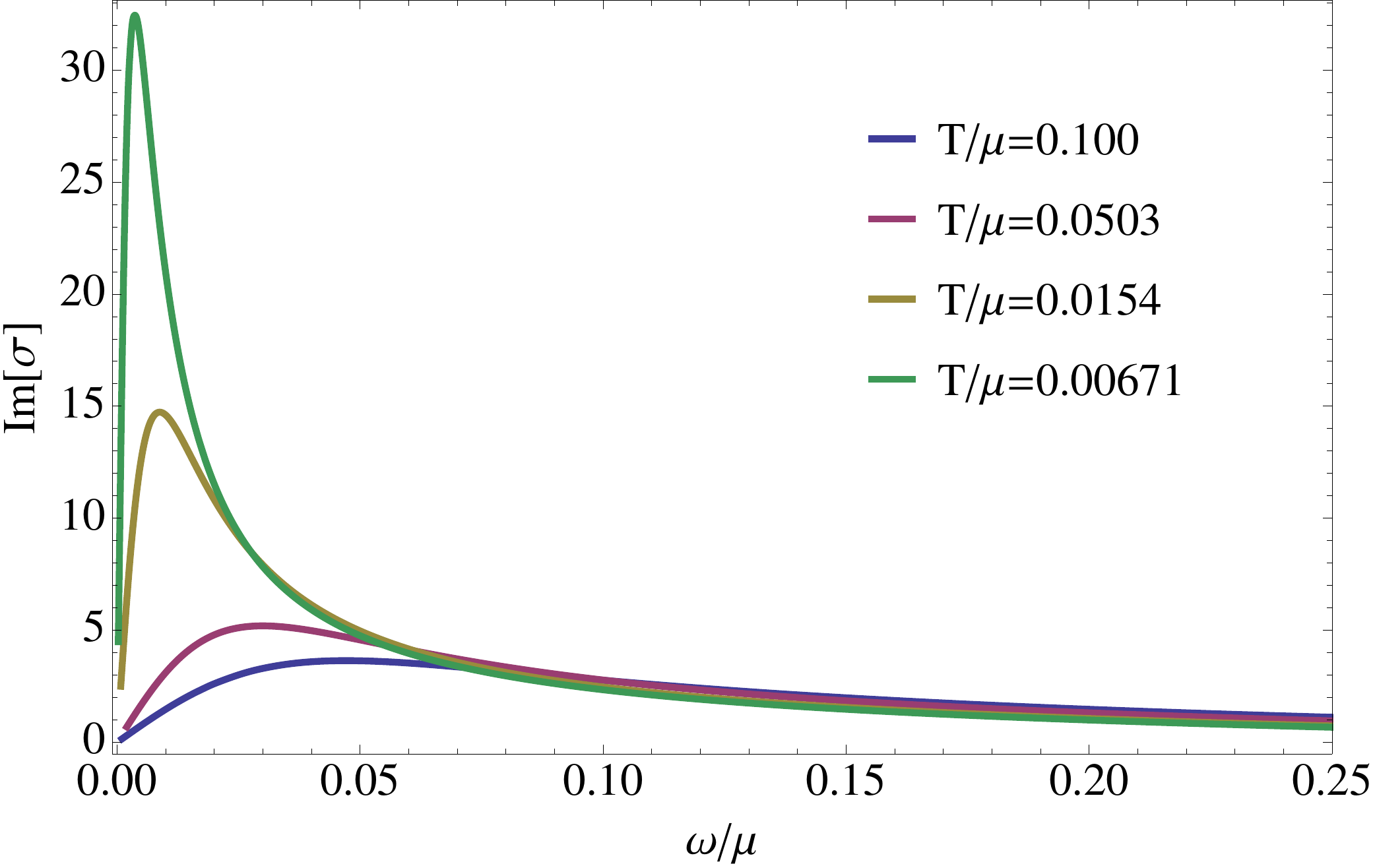}}\\
\subfloat[]{\includegraphics[width=7.3cm]{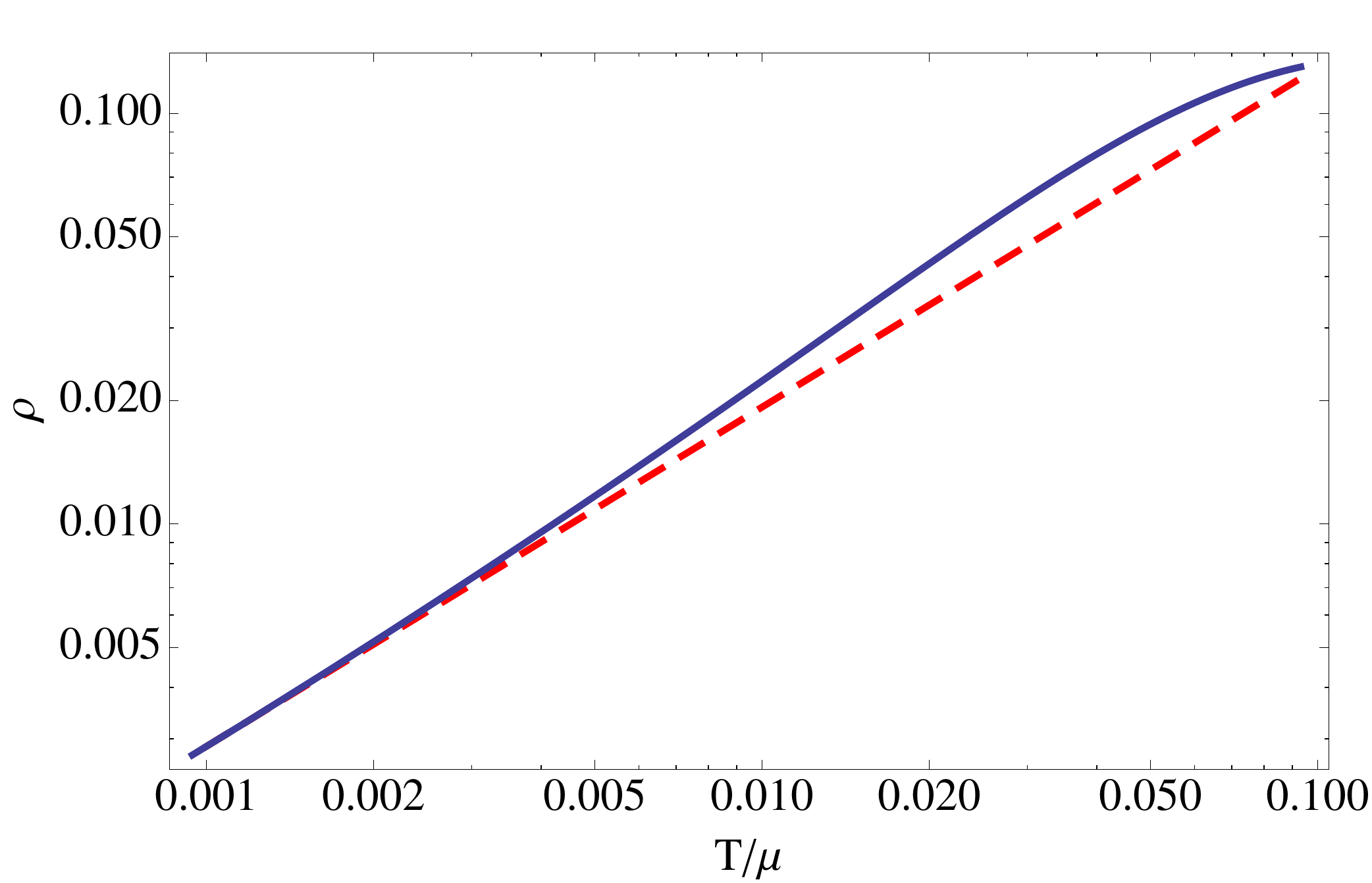}}\hskip 1 em
\subfloat[]{\includegraphics[width=7.3cm]{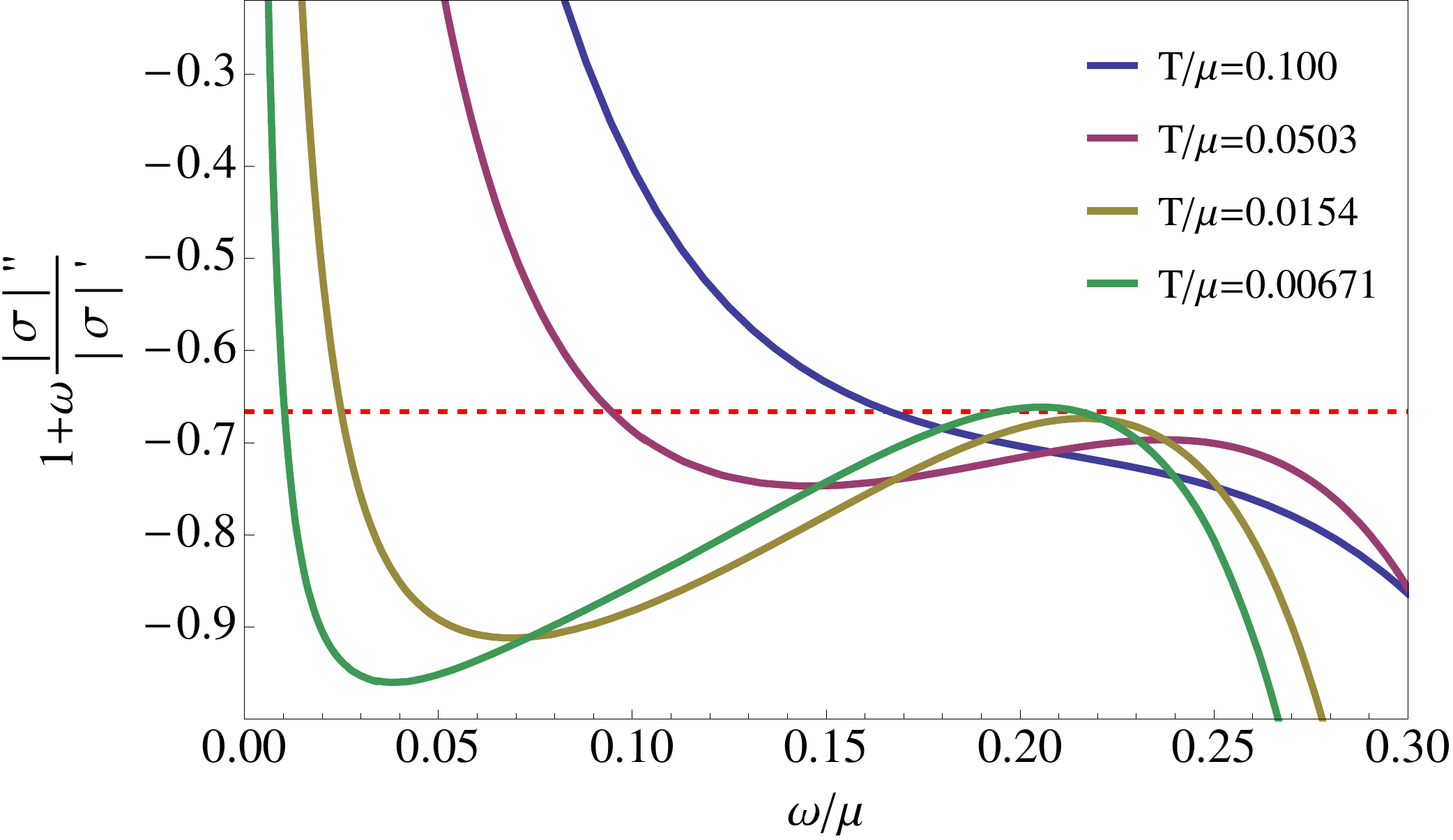}}\\
\caption{Black holes in the metallic phase for lattice parameters $\lambda/\mu=1/2$ and $k/\mu=1/\sqrt{2}$. 
Panels (a) and (b) shows the real and imaginary parts of the optical conductivity, $Re(\sigma)$ and $Im(\sigma)$, respectively, for four different temperatures. As the the temperature is lowered, the Drude peak becomes more pronounced.
Panel (c) shows the behaviour of the DC resistivity, $\rho$, as a function
of $T/\mu$. The blue line is the data and the red dashed line is the scaling expected from \eqref{dares}. 
Panel (d) shows 
a plot of $1+\omega|\sigma|''/|\sigma|'$ versus frequency; there is no evidence for an intermediate scaling of the form \eqref{scal}, 
which corresponds to the red dashed line.
}\label{figone}
\end{figure}

One might expect that this metallic scenario unfolds for large wavelength and small Q-lattice deformations of the AdS-RN black hole 
i.e. $\lambda/\mu^{3-\Delta}<<1$ and $k/\mu<<1$.
As an illustrative example, we have numerically constructed  Q-lattice black holes in the metallic phase with $\lambda/\mu=1/2$ and $k/\mu=1/\sqrt{2}$. 
By examining the properties of these solutions at very low temperatures, we find that they approach domain walls interpolating between $AdS_4$ in the UV and $AdS_2\times\mathbb{R}^2$
in the IR. In section \ref{condsec} we describe the calculation of the optical conductivity; the results
for the metallic phase black holes that we have constructed are presented in figure \ref{figone}.

In figure \ref{figone}(c) we see that the DC resistivity increases with temperature and hence we do indeed have a metallic phase.
In figures \ref{figone}(a) and \ref{figone}(b) we have plotted the real an imaginary part of the optical conductivity, respectively, for
four different temperatures. In particular, in \ref{figone}(a) we see the Drude-type peaks appearing, which get more pronounced
as the temperature is lowered. By fitting\footnote{For $\omega<<T$ we make the four parameter fit: $1/\sigma= (a_1+a_2\omega^2)-i\omega(a_3+a_4\omega^2)$, for constants $a_i$, where we used $\sigma^*(\omega)=\sigma(-\omega)$,
and we note that $a_1=(K\tau)^{-1}=\rho$ and $a_3=K^{-1}$.}
to \eqref{drude} we obtain the values for $\tau\mu$ and $K/\mu$ given in table \ref{green}.
\begin{table}[!th]
\begin{center}
\setlength{\tabcolsep}{0.45em}
\begin{tabular}{|c|c|c|c|}
\hline
$T/\mu$ & $\tau\mu$ &$K/\mu$\\
\hline
0.1 & 20 & 0.37 \\
0.0503 & 33 & 0.32 \\
0.0154 & 113 & 0.26 \\
0.00671 & 272& 0.24 \\
\hline
\end{tabular}
\end{center}
\caption{Parameters after fitting to the Drude behaviour \eqref{drude} for small $\omega$,
for the black holes in the metallic phase for lattice parameters $\lambda/\mu=1/2$ and $k/\mu=1/\sqrt{2}$. }
\label{green}
\end{table}

To observe the exact scaling behaviour $\rho\sim (T/\mu)^{2\Delta(k)-2}=(T/\mu)^{2\delta_\varphi}$, 
with $\Delta(k), \delta_\varphi$, as in \eqref{dimk}, \eqref{deltexp}, 
as predicted
by \cite{Hartnoll:2012rj}, is not straightforward because the scaling only manifests itself when $T<<\mu$. 
We have constructed the black hole solutions down to temperatures $T/\mu\sim 2.5\times 10^{-7}$
and, as noted, we find that the black holes approach the $AdS_2\times\mathbb{R}^2$ solution. By identifying $v_{10}$ with $V_{1+}$ we deduce 
that $k\sim0.707$ gets renormalised to a value $e^{-v_{10}}k\sim 2.236$ and hence
$\Delta(k)\sim 1.413$ corresponding to the scaling $\rho\sim (T/\mu)^{0.826}$.
We have calculated the conductivity for temperatures down to $T/\mu\sim 7\times10^{-4}$ and from this deduced the DC resistivity. 
The scaling behaviour eventually manifests itself at these low temperatures as one can see 
from panel (c) of figure \ref{figone}.
Our results in \ref{figone}(c) are consistent with this scaling to the order of less than 1\%.
This is the first direct check of the prediction of \cite{Hartnoll:2012rj} for back-reacted holographic 
lattices\footnote{The recent analytic results on the scaling of the DC resistivity for perturbative lattices \cite{Blake:2013owa} also confirmed
the prediction of \cite{Hartnoll:2012rj}. Note, though, that the order in perturbations that were considered do not
include back reaction of the metric and, in particular, that length scales get renormalised from the UV to the IR. Analytic results for
back-reacted Q lattice black holes will appear in \cite{Donos:2014uba}.}.
Note that for very large temperatures the resistivity should eventually approach unity, which is the constant value for the
$AdS$-Schwarzschild black hole at zero momentum \cite{Herzog:2007ij}.

We can also investigate the possibility that there is a scaling of the form \eqref{scal}, which has been reported 
for other models in the range $2\lesssim \omega\tau\lesssim8$ \cite{Horowitz:2012ky,Horowitz:2012gs,Horowitz:2013jaa,Ling:2013nxa}. 
If this scaling is present then $1+\omega|\sigma|''/|\sigma|'=-2/3$. Our results are plotted in figure \ref{figone}(d) and, for example, 
from table \ref{green} for $T/\mu=0.1$ the relevant range is $0.1\lesssim \omega/\mu\lesssim 0.4$, while for 
$T/\mu=0.00671$ it is $0.0073\lesssim \omega/\mu\lesssim 0.029$. Our results show
that there is a strong temperature dependence and there is no evidence of a mid frequency scaling region.
Note that $|\sigma|$ has a minimum at some
value of $\omega$ and hence the function $1+\omega|\sigma|''/|\sigma|'$ will diverge at that point and, furthermore for larger values of $\omega$
it will be positive. Finally we note that for very large $\omega/\mu$ and fixed $T$, the conductivity should approach that of the AdS-RN black hole with
$\sigma\to 1$ \cite{Herzog:2007ij}.

\subsection{Black holes dual to the insulating phase}

The metallic phase discussed in the last subsection arises for a given UV lattice, specified by $\lambda/\mu^{3-\Delta}$ and $k/\mu$,
whenever the $T=0$ ground state approaches $AdS_2\times\mathbb{R}^2$ in the far IR. In this section we will construct black holes
where this does not occur and we will see that they exhibit insulating behaviour. 

\begin{figure}
\centering
\subfloat[]{\includegraphics[width=7.3cm]{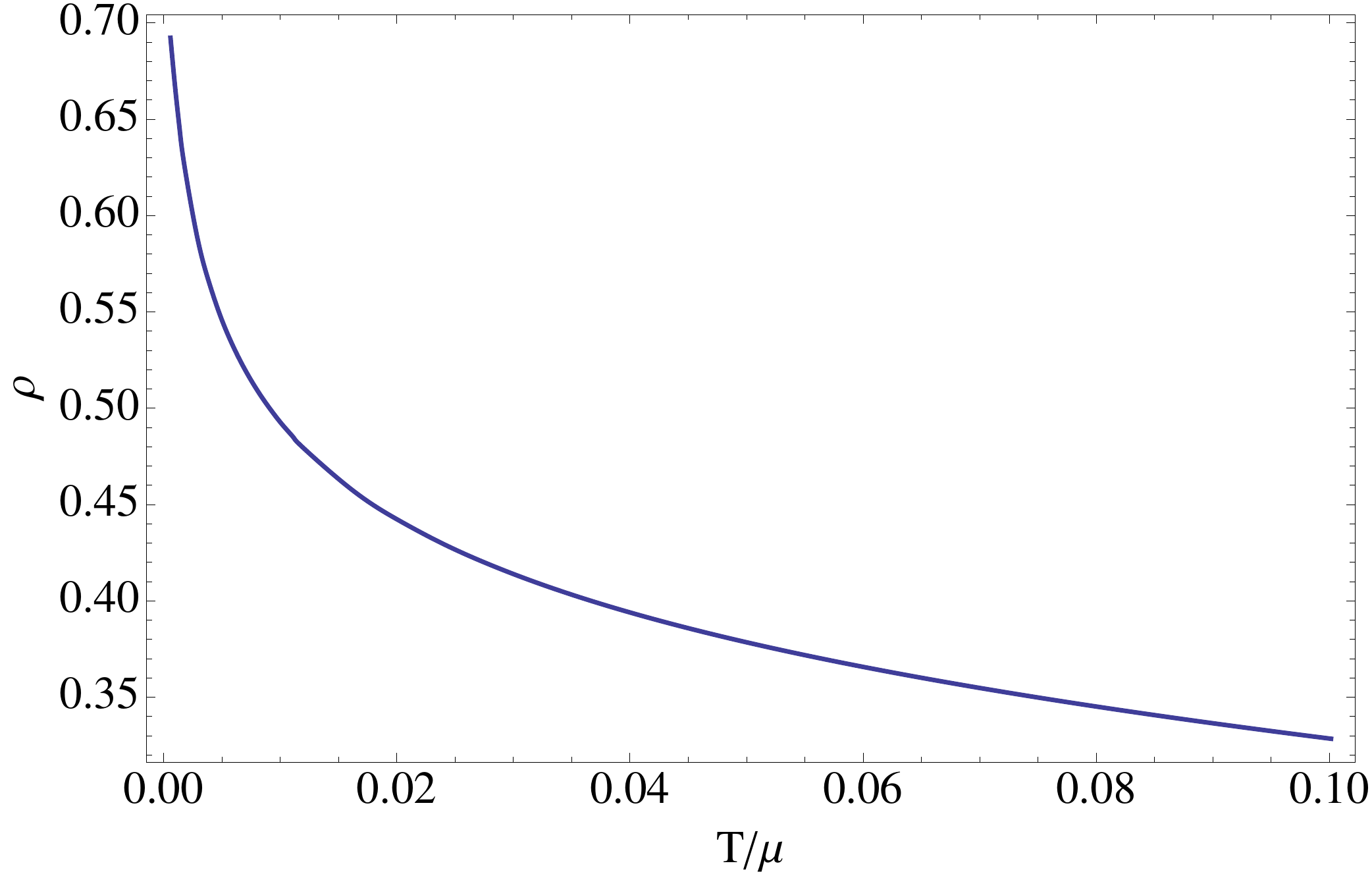}}\\
\subfloat[]{\includegraphics[width=7.3cm]{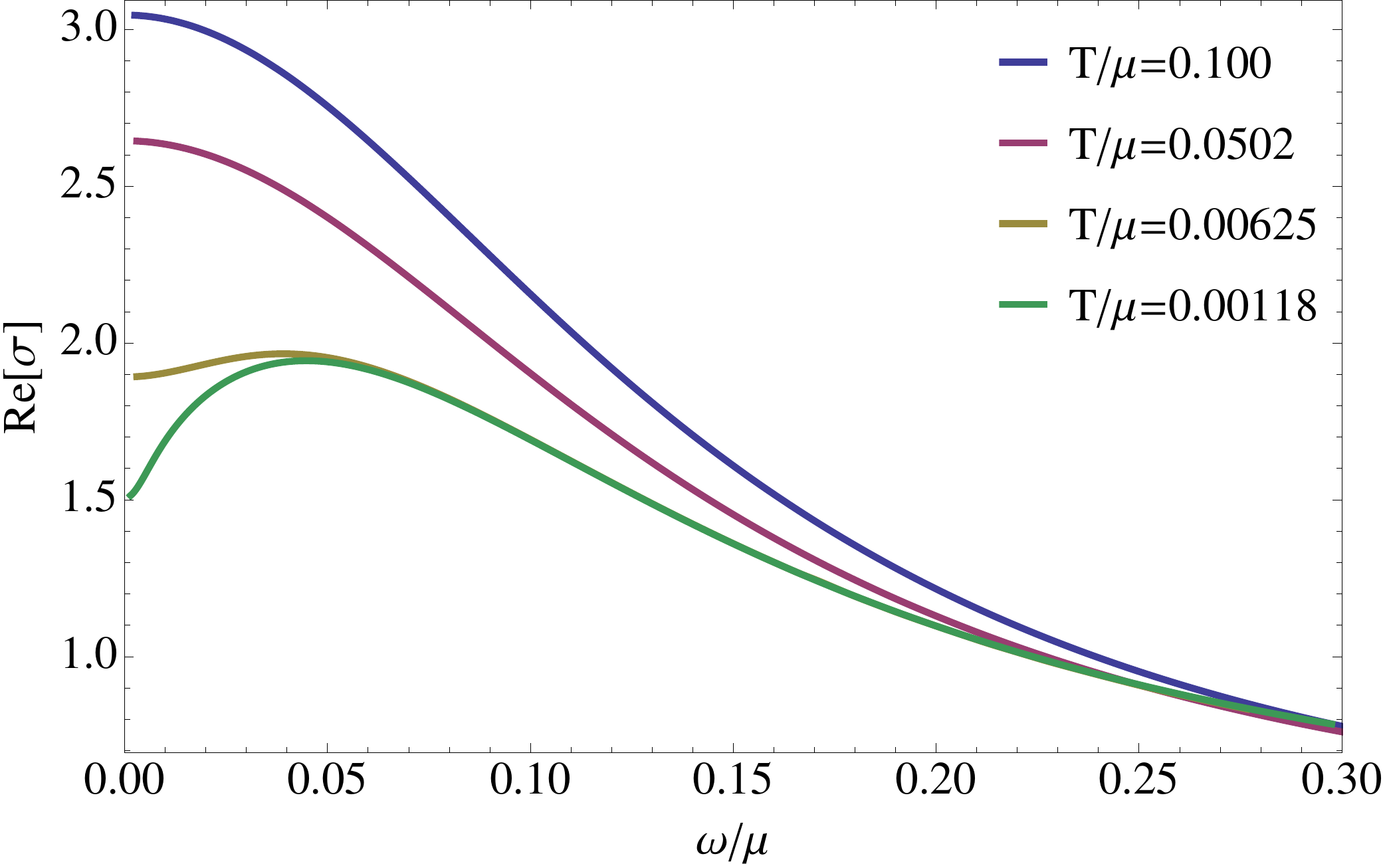}}\hskip 1 em
\subfloat[]{\includegraphics[width=7.3cm]{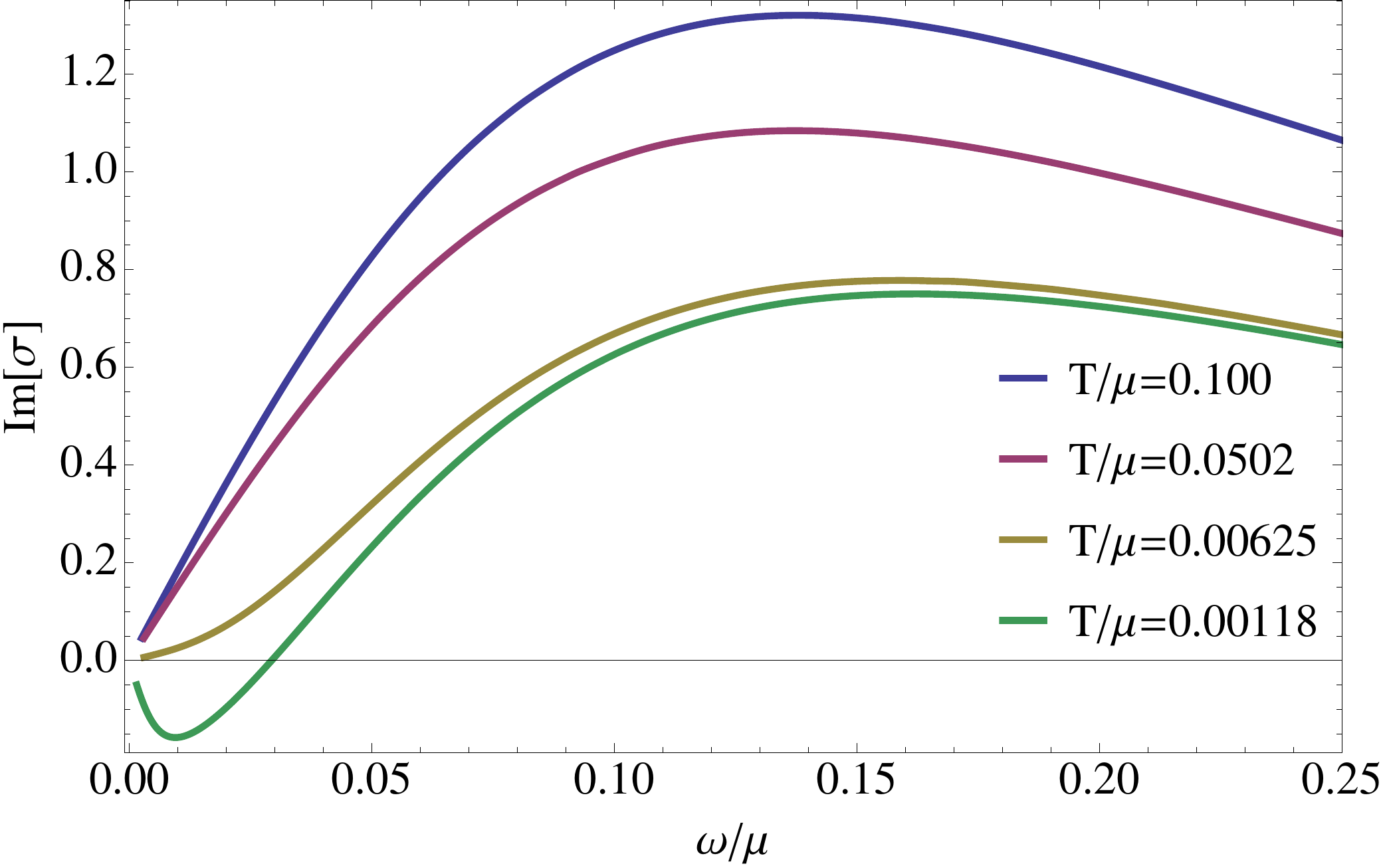}}
\caption{Black holes in the insulating phase for lattice parameters $\lambda/\mu^{3-\Delta}=2$ and $k/\mu=1/2^{3/2}$. 
Panel (a) shows the behaviour of the DC resistivity, $\rho$, as a function
of $T/\mu$. 
Panels (b) and (c) show the real and imaginary parts of the optical conductivity, $Re(\sigma)$ and $Im(\sigma)$, respectively, for four different temperatures.
For very low temperatures we see in panel (b) the suppression of spectral weight for small $\omega$ and the development of a mid-frequency hump.
}\label{figins}
\end{figure}

We focus on the specific values $\lambda/\mu^{3-\Delta}=2$ and $k/\mu=1/2^{3/2}$.
The optical conductivity and the DC resistivities for these black holes are displayed in figure \ref{figins}.
The DC resistivity is increasing as we lower the temperature indicating that the system is in an insulating phase.
Furthermore, for very low temperatures, for example $T/\mu\sim 0.00118$, we see that the real part of the
optical conductivity reveals a suppression of spectral weight for small $\omega/\mu$, with the weight being transferred to a mid frequency hump.
Very similar behaviour was seen for the helical lattice black holes dual to insulating phases in \cite{Donos:2012js}.

Lowering the temperature further we might expect to find the $T=0$ ground states for this insulating phase.
Actually this is not guaranteed as there are certainly situations in holography where black holes only
exist down to a minimum temperature, for example \cite{Donos:2011ut}. 
For the insulating black holes with the above lattice parameters we have found an interesting feature
at the low temperature $T_c/\mu\sim 2.8\times 10^{-5}$. Specifically we find that there appears to be 
a kink in the entropy density versus temperature curve, with $s'(T_c)=0$, which at first sight appears to represent a minimum temperature. However, closer detailed numerical investigation shows that there is another branch of insulating
black holes at lower temperature, with broadly similar insulating behaviour. 
The simplest interpretation is that there is a first order transition at $T_c$.
Assuming this to be the case, we have found that the low temperature branch exists at least down to the ultra low temperatures
$T_c/\mu\sim 10^{-9}$. Furthermore, we find that the entropy density is going to zero and that the solutions are becoming singular.
We are particularly interested in extracting the far IR behaviour of
the $T=0$ black holes. However, in general, this is a non-trivial task unless some simplification represents
itself in the numerical solutions, such as the functions approaching a power-law behaviour. We have not been
able to find any evidence for such power law behaviour in the present setting.

It would be certainly interesting to explore these issues further. 
Note that we have considered other values for the UV lattice data, finding
somewhat similar results, but a more comprehensive analysis of the behaviour for general values of $\lambda/\mu^{3-\Delta}$ and $k/\mu$ is
left for future work. One point that is worth highlighting is that the model also possesses another fixed point solution that may play an
important role in understanding the phase structure of the model. As we describe in the appendix there is
a novel electrically neutral $AdS_3\times\mathbb{R}$ fixed point solution with a spectrum containing modes corresponding to
both irrelevant and relevant operators. The presence of the relevant operator indicates that for generic lattice data 
it will not be possible to construct domain wall solutions interpolating between $AdS_4$ in the UV and $AdS_3\times\mathbb{R}$ in the IR. However, it is possible
that a fine tuned domain wall solution exists for specific lattice data, which might correspond to an unstable RG flow providing a bifurcation
between the metallic and insulating behaviours analogous to what was observed for the helical black hole lattices in \cite{Donos:2011ut}.

\section{Conductivity}\label{condsec}
In this section we explain how we calculate the conductivity for the black holes that we have constructed.
Although the general idea is standard, the technical implementation in the presence of the lattice deformation
warrants some discussion.
We consider the following consistent linear perturbation about the black hole solutions
\begin{align}\label{eq:pert_ansatz1}
\delta g_{tx_1}&= \delta h_{tx_1}(t,r)\,,\nn
\delta A_{x_1}&=\delta a_{x_1}(t,r)\,,\nn
\delta\phi&=ie^{ikx_1}\delta\varphi(t,r)\,,
\end{align}
where $\delta h_{tx_1}, \delta a_{x_1}$ and $\delta\varphi$ are all {\it real} functions of $(t,r)$ and we note the factor of $i$ in the last line.
After substituting into the equations of motion we obtain real partial differential equations.
We next allow for a time dependence of the form $e^{-i\omega t}$ by writing
\begin{align}\label{eq:pert_ansatz2}
\delta h_{tx_1}(t,r)&=e^{-i\omega t} \delta h_{tx_1}(r)\,,\nn
\delta a_{x_1}(t,r)&=e^{-i\omega t}\delta a_{x_1}(r)\,,\nn
\delta\varphi(t,r)&=e^{-i\omega t}\delta\varphi(r)\,,
\end{align}
and we are lead to the following system of ODEs:
\begin{align}\label{eq:pert_eom}
&\delta a_{x_1}''+\left(U^{-2}\omega^2-U^{-1}a'^2\right)\delta a_{x_1}+\left(U^{-1}U'-V_1'+V_2'\right)\delta a_{x_1}'
+2i\frac{k}{\omega}a'\left(\varphi'\delta\varphi-\varphi\delta \varphi'\right)=0,\nn
&\delta\varphi''
+\left(U^{-2}\omega^2-m^2U^{-1}-k^2 U^{-1}e^{-2 V_1}\right)\delta\varphi
+\left(U^{-1}U'+V_1'+V_2'\right)\delta\varphi'
-i k \omega U^{-2} e^{-2 V_1} \varphi \delta h_{tx_1}=0\,,\nn
&\delta h_{tx_1}'+ a' \delta a_{x_1} -2   V_1'\delta h_{tx_1} -2 i \frac{k}{\omega} U \left(\varphi'\delta\varphi -\varphi\delta\varphi'\right)=0\,.
\end{align}

At the black hole event horizon we impose purely ingoing boundary conditions with the perturbations behaving as
\begin{align}\label{irexp}
\delta a_{x_1}&=(r-r_+)^{-i\omega/4\pi T}\left(\delta a_{x_1}^{(+)}+\dots \right)\,,\nn
\delta\varphi&=(r-r_+)^{-i\omega/4\pi T}\left(\delta\varphi^{(+)}+\dots \right)\,,\nn
 \delta h_{tx_1}&=(r-r_+)^{-i\omega/4\pi T}(\delta h_{tx_1}^{(+)}(r-r_+)+\dots)\,,
\end{align}
where the dots refer to terms higher order in $(r-r_+)$.
The regularity of this perturbation at the black horizon can be seen by using ingoing Eddington-Finklestein
coordinates $(v,r)$ with $v=t+\log(r-r_+)^{4\pi T}$. Using the equations of motion we find that this
expansion is fixed by two parameters $\delta a_{x_1}^{(+)}$, $\delta\varphi^{(+)}$ with 
\begin{align}
\delta h_{tx_1}^{(+)}=-\frac{a_+\delta a_{x_1}^{(+)}+2k\varphi_+\delta\varphi^{(+)}}{r_+^2(1-i\frac{\omega}{4\pi T})}\,.
\end{align}

In the UV we impose that as $r\to\infty$:
\begin{align}\label{uvex}
\delta h_{tx_1}&={\delta h_{tx_1}^{(0)}}{r}^2+\dots\,,\nn
\delta a_{x_1}&=\delta a_{x_1}^{(0)}+\frac{\delta a_{x_1}^{(1)}}{r}+\dots\,,\nn
\delta\varphi&=\frac{\delta \varphi^{(0)}}{r^{3-\Delta}}+\dots +\frac{\delta \varphi^{(1)}}{r^{\Delta} }+\dots\,.\,.
\end{align}
Now we are interested in a perturbation that switches on an electric field and then we want to read off the current to
obtain the conductivity. One might be tempted to set $\delta h_{tx_1}^{(0)}=\delta \varphi^{(0)}=0$ but this over constrains the system.
To see this we note that 
a solution to the ODEs \eqref{eq:pert_eom} is specified by five parameters. From the IR and UV expansions \eqref{irexp}, \eqref{uvex}
we have a total of seven parameters. However, since
the ODEs \eqref{eq:pert_eom} are linear we can scale one of the seven parameters to unity, leaving six. This means that we need to
impose just one more constraint on the parameters. This constraint can be found as follows. 

To ensure that we are extracting just the current-current correlator, 
we can use diffeomorphisms and gauge-transformations to demand that the perturbation satisfies, as $r\to\infty$,
\begin{align}
\frac{1}{r^2}\left(\delta g_{\mu\nu}+{\cal L}_\zeta g_{\mu\nu}\right)&\to0\,,\nn
\delta A+{\cal L}_\zeta A+d\Lambda&\to e^{-i\omega t} \mu_{x_1}dx_1\,,\nn
r^{3-\Delta}\left(\delta\phi+{\cal L}_\zeta \phi\right)&\to 0\,,
\end{align}
where $\zeta^\mu$ and $\Lambda$ are smooth and $\mu_{x_1}$ will be the source for the current.
For our specific set-up we can take $\Lambda=0$ and the only non-vanishing component of $\zeta^\mu$ to be
$\zeta^x=\epsilon e^{-i\omega t}$ where
$\epsilon$ is a small parameter. From this we can deduce that we have $\mu_{x_1}=\delta a_{x_1}^{(0)}$ and that we
should impose the condition
\begin{align}\label{constp} 
\delta \varphi^{(0)}-i\frac{k\lambda}{\omega}\delta h_{tx_1}^{(0)}=0\,.
\end{align}
The optical conductivity is then given by
\begin{align}
\sigma(\omega)=-\frac{i}{\omega}\frac{\delta a_{x_1}^{(1)}}{\delta a_{x_1}^{(0)}}\,.
\end{align}
The DC resistivity is given by $\rho=1/\sigma(0)$. It is worth mentioning that 
to calculate $\rho$ numerically, one needs to calculate the optical conductivity for $\omega<<T$.

\section{Final comments}\label{fincom}

We have studied holographic Q-lattices for Einstein-Maxwell theory coupled to a single complex scalar field in $D=4$ space-time dimensions. 
We have shown that the system exhibits both metallic and insulating phases. The metallic phase is governed by the 
electrically charged $AdS_2\times\mathbb{R}^2$ solution that appears in the IR region of the $T=0$ electrically charged AdS-RN solution. 
We showed in detail that the phase exhibits a Drude-type peak and furthermore, 
at low temperatures the DC resistivity exhibits a scaling behaviour confirming the prediction
of \cite{Hartnoll:2012rj}.

We have also constructed Q-lattice black holes in a new
insulating phase down to very low temperatures. 
For temperatures lower than $T/\mu\sim 10^{-3}$ we see a transferral of spectral weight in the optical conductivity and the generation of a mid frequency hump. At temperatures $T/\mu\sim 2.8\times10^{-5}$ we have found evidence for a first order transition to another branch of insulating black holes. It would
be interesting to investigate these further including trying to elucidate the ultimate IR ground states at $T=0$ which seem to have vanishing entropy density.
A possibly related issue, is to further
understand the role played by the neutral $AdS_3\times\mathbb{R}$ ground state that we have found and discussed in the appendix.

We focussed on the case where the mass of the complex scalar is given by $m^2=-3/2$ with $\Delta=(3+\sqrt{3})/2$ in the $d=3$ CFT,
which saturates the $AdS_2\times\mathbb{R}^2$ BF bound, corresponding to a stable metallic phase.  
We have also made some numerical investigations into the case $m^2=-2$ with $\Delta=2$ in the $d=3$ CFT. 
We have constructed black holes
with conductivities exhibiting metallic and insulating behaviours much as in figure \ref{figone}.
However, for this case the complex scalar violates the $AdS_2\times\mathbb{R}^2$ BF bound
and hence, at least for the metallic black holes, one will find an additional new phase appearing
at low 
temperatures\footnote{The same is true for the model considered in \cite{Horowitz:2012ky}.}. 
When there is no lattice deformation a possible ground state for this model was identified in \cite{Horowitz:2009ij}.
It will be interesting to see how this is modified by the lattice deformation and also to investigate the impact on the
insulating phase.

It is also natural to consider a more general class of models including a coupling of the scalar field to the gauge field and
a more general potential than the simple mass term. We expect that within this more general class of models it will
be possible to obtain the many novel IR ground states in explicit form \cite{Donos:2014uba}. 
It will be particularly interesting to explore interconnections with charge density waves \cite{Donos:2013gda} which should lead to close analogues of Mott insulating ground states. Such models can be studied in various spacetime dimensions.

For the Q-lattices that we have constructed for specific values of lattice strength $\lambda$ and wave-number $k$, for both $m^2=-3/2$ and $m^2=-2$, 
we find no evidence that the metallic phase has an intermediate scaling of the form \eqref{scal}. 
How can this be reconciled with the results reported in 
\cite{Horowitz:2012gs,Horowitz:2013jaa,Horowitz:2012ky,Ling:2013nxa}, where numerical evidence for this behaviour was found and 
moreover it was suggested that this might be a universal feature of holographic lattices? 
One possibility is that the numerical evidence found in those papers is actually misleading and in fact there is not a robust 
power-law behaviour for the lattices considered.

An interesting perspective is to consider the same model \eqref{act}
that we have in this paper, but with a family of
lattice deformations, labelled by $\alpha$, given by
\begin{align}
\phi=\sqrt{2}\lambda\left(\cos\alpha\cos k x_1+i \sin\alpha\sin kx_1\right)\frac{1}{r^{3-\Delta}}+\dots
\end{align}
as $r\to\infty$. For $\alpha=\pi/4$ this gives the family Q-lattices that we discussed in this paper, while for $\alpha=0$ it gives the lattices discussed in  
\cite{Horowitz:2012gs} (who just considered $m^2=-2$). 
Notice that the strength of the lattice, $\lambda$, does not depend on $\alpha$ and also that for $\alpha\ne (2n+1)\pi/4$, for 
integer $n$, the metric
will be co-homogeneity two and 
one will need to solve PDEs. 

For this general family of lattices we can use the results of \cite{Hartnoll:2012rj} and also
of \cite{Edalati:2010pn,Donos:2013gda} to deduce the scaling behaviour of the DC resistivity in the metallic phase. 
In addition to the scalar mode with wave-number $k$,
with dimension \eqref{dimk} in the IR, one also needs to take into account\footnote{Note that there will also be scalar modes with wave-number $nk$ and longitudinal modes with wave-number $2nk$, for $n>1$, but these will be more irrelevant in the IR and hence will not dominate the scaling of the DC resistivity.} 
longitudinal modes involving perturbations in
$A_t, A_{x_1}$ and $g_{tt}, g_{x_1x_1},g_{tx_1}, g_{x_2x_2}$
and with wave-number $2k$ (corresponding to the fact that the scalar lattice sources them at least at quadratic order). From 
the analysis presented in \cite{Donos:2013gda} (in particular equation (2.17)), one can deduce that when $m^2\le -1/4$ 
the DC resistivity scaling will always be determined by the decoupled scalar mode in the IR. 
Interestingly for $-1/4<m^2< 0$, for certain windows of $k$, the scaling can be determined by the longitudinal modes.
Note in particular, for the scalar lattice in \cite{Horowitz:2012ky} with $m^2=-2$ and $\alpha=0$, we are arguing that the 
DC resistivity scaling is actually 
governed by the scalar mode and not one of  the longitudinal modes as was 
stated in \cite{Horowitz:2012ky}.
Note that this work also claimed to see a numerical fit to a scaling governed by
the longitudinal mode: we believe that the fitting was misleading and that continuing to lower temperatures will 
reveal the scaling behaviour that we are predicting.

It is also worth pointing out that we do not expect the black hole solutions will be substantially different as we vary $\alpha$ away from $\pi/4$, despite the fact that one is solving PDEs instead of ODEs as in this paper.
While additional harmonics of the bulk fields will play a role, the higher harmonics are expected to be exponentially suppressed. In fact this was
seen in the numerical work in \cite{Horowitz:2012ky}. Thus it is natural to expect that conductivity for non-zero $\omega$ is also not substantially 
different from what we have seen in this paper.

All of the constructions in this paper have just involved classical gravity. it is worth recalling, however, that there 
are good reasons to expect that there are no global symmetries in theories of quantum gravity (e.g. \cite{Banks:2010zn}). 
One point of view is that we are just studying a sector of a larger classical theory that does not have a global symmetry.
Alternatively we can view the breaking of the continuous symmetry as a higher order effect in the large $N$ expansion. 
Within these contexts, or closely related ones, we think that top-down constructions should be possible.

Finally we point out that the holographic lattice constructions that we have discussed in this paper, where the translation symmetry is broken 
explicitly, can also be adapted to situations where the the symmetry is broken spontaneously.

\section*{Acknowledgements}
We thank Paul Chesler, Sean Hartnoll, Diego Hofman, Elias Kiritsis, Da-Wei Pang, Jorge Santos, Julian Sonner, David Tong and David Vegh for helpful conversations.
The work is supported by STFC grant ST/J0003533/1 and also by the European Research Council under the European Union's Seventh Framework Programme (FP7/2007-2013), ERC Grant agreement STG 279943, ``Strongly Coupled Systems".

\appendix
\section{A novel $AdS_3\times\mathbb{R}$ solution}
Provided that $m^2<0$ (equivalently, the operator dual to $\phi$ in the $d=3$ CFT dual to the $AdS_4$ vacuum is a relevant operator),
the model \eqref{act} admits an electrically neutral $AdS_3\times \mathbb{R}$ solution given by
\begin{align}
ds^2&=\frac{1}{3}ds^2(AdS_3)+dx_1^2\,,\nn
\phi&=\frac{6}{-m^2}e^{i\sqrt{-m^2}x_1}\,,
\end{align}
with $A=0$.

To explore whether there are domain wall solutions which can connect this solution with $AdS_4$, we investigate the spectrum
for this fixed point.
Within our ansatz \eqref{ansatzbh} we can consider the perturbations given by
\begin{align}\label{exd}
U&=3r^2(1+u_1 r^\delta),\qquad V_1=v_{11}r^\delta,\qquad V_2=\log(r)+v_{21}r^\delta,\nn 
a&=a_1 r^{1+\delta},\qquad\phi=\left(\frac{6}{-m^2}\right)^{1/2}e^{i\sqrt{-m^2}x_1}\phi_1 r^\delta\,.
\end{align}
These perturbations correspond to scaling dimension $\Delta=-\delta$ or $\Delta=\delta+2$
in the $d=2 $ CFT dual to the $AdS_3\times\mathbb{R}$
solution.
We find that the exponents come in four pairs with $\delta_++\delta_-=-2$ and there is an unpaired mode with $\delta=-1$. 
The paired modes have $\delta_+$ values given by $0,-1$
and
\begin{align}\label{vals}
\delta_1=-1+\frac{1}{\sqrt{3}}\sqrt{9-2\sqrt{3}\sqrt{3-m^2}},\qquad \delta_2=-1+\frac{1}{\sqrt{3}}\sqrt{9+2\sqrt{3}\sqrt{3-m^2}}\,.
\end{align}
We see that in the mass range $-9/4\le m^2<0$, which is relevant for trying to map onto $AdS_4$ in the UV, 
$\delta_1$ corresponds to a relevant operator (i.e. $\delta_1<0$) and $\delta_2$ corresponds to an irrelevant operator (i.e. $\delta_2>0$).
Note that both of these deformations have $a_1=0$ in \eqref{exd} and do not involve the gauge-field.

A parameter count now reveals that, generically, because of the presence of the relevant operator,
there will not be domain wall solutions interpolating between the lattice deformed 
$AdS_4$ in the UV and $AdS_3\times\mathbb{R}$ in the IR. However, there is the possibility that there is a fine-tuned domain wall
solution. If this exists it might correspond to a bifurcating, unstable RG solution, separating the metallic and insulating behaviours, as
in figure 2 of \cite{Donos:2012js}.

More generally, we expect that there are closely related models where the $AdS_3\times\mathbb{R}$ geometry has irrelevant operators in the IR so
that one can construct domain walls that interpolate from the Q-lattice deformed $AdS_4$ in the UV. Furthermore, changing the dimension of space-time
and the number, $n$, of spatial directions where translation invariance is broken by the holographic Q-lattice will allow one to construct domain walls from
$AdS_D$ in the UV and various $AdS_{D-n}\times\mathbb{R}^n$ in the IR. This will be explored in detail elsewhere.

\appendix
\bibliographystyle{utphys}
\bibliography{helical}{}

\providecommand{\href}[2]{#2}\begingroup\raggedright\begin{thebibliography}{10}

\bibitem{Horowitz:2012ky}
G.~T. Horowitz, J.~E. Santos, and D.~Tong, ``{Optical Conductivity with
  Holographic Lattices},''
  \href{http://dx.doi.org/10.1007/JHEP07(2012)168}{{\em JHEP} {\bfseries 1207}
  (2012) 168},
\href{http://arxiv.org/abs/1204.0519}{{\ttfamily arXiv:1204.0519 [hep-th]}}.

\bibitem{Horowitz:2012gs}
G.~T. Horowitz, J.~E. Santos, and D.~Tong, ``{Further Evidence for
  Lattice-Induced Scaling},''
  \href{http://dx.doi.org/10.1007/JHEP11(2012)102}{{\em JHEP} {\bfseries 1211}
  (2012) 102},
\href{http://arxiv.org/abs/1209.1098}{{\ttfamily arXiv:1209.1098 [hep-th]}}.

\bibitem{Horowitz:2013jaa}
G.~T. Horowitz and J.~E. Santos, ``{General Relativity and the Cuprates},''
\href{http://arxiv.org/abs/1302.6586}{{\ttfamily arXiv:1302.6586 [hep-th]}}.

\bibitem{Donos:2012js}
A.~Donos and S.~A. Hartnoll, ``{Interaction-driven localization in
  holography},'' \href{http://dx.doi.org/10.1038/nphys2701}{{\em Nature Phys.}
  {\bfseries 9} (2013) 649--655},
\href{http://arxiv.org/abs/1212.2998}{{\ttfamily arXiv:1212.2998}}.

\bibitem{Ling:2013nxa}
Y.~Ling, C.~Niu, J.-P. Wu, and Z.-Y. Xian, ``{Holographic Lattice in
  Einstein-Maxwell-Dilaton Gravity},''
  \href{http://dx.doi.org/10.1007/JHEP11(2013)006}{{\em JHEP} {\bfseries 1311}
  (2013) 006},
\href{http://arxiv.org/abs/1309.4580}{{\ttfamily arXiv:1309.4580 [hep-th]}}.

\bibitem{Chesler:2013qla}
P.~Chesler, A.~Lucas, and S.~Sachdev, ``{Conformal field theories in a periodic
  potential: results from holography and field theory},''
  \href{http://dx.doi.org/10.1103/PhysRevD.89.026005}{{\em Phys.Rev.}
  {\bfseries D89} (2014) 026005},
\href{http://arxiv.org/abs/1308.0329}{{\ttfamily arXiv:1308.0329 [hep-th]}}.

\bibitem{Karch:2007pd}
A.~Karch and A.~O'Bannon, ``{Metallic AdS/CFT},''
  \href{http://dx.doi.org/10.1088/1126-6708/2007/09/024}{{\em JHEP} {\bfseries
  0709} (2007) 024},
\href{http://arxiv.org/abs/0705.3870}{{\ttfamily arXiv:0705.3870 [hep-th]}}.

\bibitem{Hartnoll:2007ih}
S.~A. Hartnoll, P.~K. Kovtun, M.~Muller, and S.~Sachdev, ``{Theory of the
  Nernst effect near quantum phase transitions in condensed matter, and in
  dyonic black holes},''
  \href{http://dx.doi.org/10.1103/PhysRevB.76.144502}{{\em Phys. Rev.}
  {\bfseries B76} (2007) 144502},
\href{http://arxiv.org/abs/0706.3215}{{\ttfamily arXiv:0706.3215
  [cond-mat.str-el]}}.

\bibitem{Hartnoll:2007ip}
S.~A. Hartnoll and C.~P. Herzog, ``{Ohm's Law at strong coupling: S duality and
  the cyclotron resonance},''
  \href{http://dx.doi.org/10.1103/PhysRevD.76.106012}{{\em Phys. Rev.}
  {\bfseries D76} (2007) 106012},
\href{http://arxiv.org/abs/0706.3228}{{\ttfamily arXiv:0706.3228 [hep-th]}}.

\bibitem{Hartnoll:2009ns}
S.~A. Hartnoll, J.~Polchinski, E.~Silverstein, and D.~Tong, ``{Towards strange
  metallic holography},'' \href{http://dx.doi.org/10.1007/JHEP04(2010)120}{{\em
  JHEP} {\bfseries 1004} (2010) 120},
\href{http://arxiv.org/abs/0912.1061}{{\ttfamily arXiv:0912.1061 [hep-th]}}.

\bibitem{Faulkner:2010zz}
T.~Faulkner, N.~Iqbal, H.~Liu, J.~McGreevy, and D.~Vegh, ``{Strange metal
  transport realized by gauge/gravity duality},''
\href{http://dx.doi.org/10.1126/science.1189134}{{\em Science} {\bfseries 329}
  (2010) 1043--1047}.

\bibitem{Hartnoll:2012rj}
S.~A. Hartnoll and D.~M. Hofman, ``{Locally Critical Resistivities from Umklapp
  Scattering},'' \href{http://dx.doi.org/10.1103/PhysRevLett.108.241601}{{\em
  Phys.Rev.Lett.} {\bfseries 108} (2012) 241601},
\href{http://arxiv.org/abs/1201.3917}{{\ttfamily arXiv:1201.3917 [hep-th]}}.

\bibitem{Liu:2012tr}
Y.~Liu, K.~Schalm, Y.-W. Sun, and J.~Zaanen, ``{Lattice Potentials and Fermions
  in Holographic non Fermi-Liquids: Hybridizing Local Quantum Criticality},''
  \href{http://dx.doi.org/10.1007/JHEP10(2012)036}{{\em JHEP} {\bfseries 1210}
  (2012) 036},
\href{http://arxiv.org/abs/1205.5227}{{\ttfamily arXiv:1205.5227 [hep-th]}}.

\bibitem{Vegh:2013sk}
D.~Vegh, ``{Holography without translational symmetry},''
\href{http://arxiv.org/abs/1301.0537}{{\ttfamily arXiv:1301.0537 [hep-th]}}.

\bibitem{Mahajan:2013cja}
R.~Mahajan, M.~Barkeshli, and S.~A. Hartnoll, ``{Non-Fermi liquids and the
  Wiedemann-Franz law},''
  \href{http://dx.doi.org/10.1103/PhysRevB.88.125107}{{\em Phys.Rev.}
  {\bfseries B88} (2013) 125107},
\href{http://arxiv.org/abs/1304.4249}{{\ttfamily arXiv:1304.4249
  [cond-mat.str-el]}}.

\bibitem{Davison:2013jba}
R.~A. Davison, ``{Momentum relaxation in holographic massive gravity},''
  \href{http://dx.doi.org/10.1103/PhysRevD.88.086003}{{\em Phys.Rev.}
  {\bfseries D88} (2013) 086003},
\href{http://arxiv.org/abs/1306.5792}{{\ttfamily arXiv:1306.5792 [hep-th]}}.

\bibitem{Faulkner:2013bna}
T.~Faulkner, N.~Iqbal, H.~Liu, J.~McGreevy, and D.~Vegh, ``{Charge transport by
  holographic Fermi surfaces},''
  \href{http://dx.doi.org/10.1103/PhysRevD.88.045016}{{\em Phys.Rev.}
  {\bfseries D88} (2013) 045016},
\href{http://arxiv.org/abs/1306.6396}{{\ttfamily arXiv:1306.6396 [hep-th]}}.

\bibitem{2003Natur.425..271M}
D.~v.~d. {Marel}, H.~J.~A. {Molegraaf}, J.~{Zaanen}, Z.~{Nussinov},
  F.~{Carbone}, A.~{Damascelli}, H.~{Eisaki}, M.~{Greven}, P.~H. {Kes}, and
  M.~{Li}, ``{Quantum critical behaviour in a high-T$_{c}$ superconductor},''
  \href{http://dx.doi.org/10.1038/nature01978}{{\em Nature} {\bfseries 425}
  (Sept., 2003) 271--274},
  \href{http://arxiv.org/abs/arXiv:cond-mat/0309172}{{\ttfamily
  arXiv:cond-mat/0309172}}.

\bibitem{2006AnPhy.321.1716V}
D.~{van der Marel}, F.~{Carbone}, A.~B. {Kuzmenko}, and E.~{Giannini},
  ``{Scaling properties of the optical conductivity of Bi-based cuprates},''
  \href{http://dx.doi.org/10.1016/j.aop.2006.04.012}{{\em Annals of Physics}
  {\bfseries 321} (July, 2006) 1716--1729},
  \href{http://arxiv.org/abs/arXiv:cond-mat/0604037}{{\ttfamily
  arXiv:cond-mat/0604037}}.

\bibitem{Blake:2013owa}
M.~Blake, D.~Tong, and D.~Vegh, ``{Holographic Lattices Give the Graviton a
  Mass},''
\href{http://arxiv.org/abs/1310.3832}{{\ttfamily arXiv:1310.3832 [hep-th]}}.

\bibitem{Coleman:1985ki}
S.~R. Coleman, ``{Q Balls},''
\href{http://dx.doi.org/10.1016/0550-3213(85)90286-X}{{\em Nucl.Phys.}
  {\bfseries B262} (1985) 263}.

\bibitem{Amado:2013xya}
I.~Amado, D.~Arean, A.~Jimenez-Alba, K.~Landsteiner, L.~Melgar, {\em et al.},
  ``{Holographic Type II Goldstone bosons},''
  \href{http://dx.doi.org/10.1007/JHEP07(2013)108}{{\em JHEP} {\bfseries 1307}
  (2013) 108},
\href{http://arxiv.org/abs/1302.5641}{{\ttfamily arXiv:1302.5641 [hep-th]}}.

\bibitem{Donos:2014uba}
A.~Donos and J.~P. Gauntlett, ``{Novel metals and insulators from
  holography},''
\href{http://arxiv.org/abs/1401.5077}{{\ttfamily arXiv:1401.5077 [hep-th]}}.

\bibitem{Herzog:2007ij}
C.~P. Herzog, P.~Kovtun, S.~Sachdev, and D.~T. Son, ``{Quantum critical
  transport, duality, and M-theory},''
  \href{http://dx.doi.org/10.1103/PhysRevD.75.085020}{{\em Phys.Rev.}
  {\bfseries D75} (2007) 085020},
\href{http://arxiv.org/abs/hep-th/0701036}{{\ttfamily arXiv:hep-th/0701036
  [hep-th]}}.

\bibitem{Donos:2011ut}
A.~Donos and J.~P. Gauntlett, ``{Superfluid black branes in $AdS_4\times
  S^7$},'' \href{http://dx.doi.org/10.1007/JHEP06(2011)053}{{\em JHEP}
  {\bfseries 06} (2011) 053},
\href{http://arxiv.org/abs/1104.4478}{{\ttfamily arXiv:1104.4478 [hep-th]}}.

\bibitem{Horowitz:2009ij}
G.~T. Horowitz and M.~M. Roberts, ``{Zero Temperature Limit of Holographic
  Superconductors},''
  \href{http://dx.doi.org/10.1088/1126-6708/2009/11/015}{{\em JHEP} {\bfseries
  0911} (2009) 015},
\href{http://arxiv.org/abs/0908.3677}{{\ttfamily arXiv:0908.3677 [hep-th]}}.

\bibitem{Donos:2013gda}
A.~Donos and J.~P. Gauntlett, ``{Holographic charge density waves},'' {\em
  Phys.Rev.} {\bfseries D87} (2013) 126008,
\href{http://arxiv.org/abs/1303.4398}{{\ttfamily arXiv:1303.4398 [hep-th]}}.

\bibitem{Edalati:2010pn}
M.~Edalati, J.~I. Jottar, and R.~G. Leigh, ``{Holography and the sound of
  criticality},'' \href{http://dx.doi.org/10.1007/JHEP10(2010)058}{{\em JHEP}
  {\bfseries 1010} (2010) 058},
\href{http://arxiv.org/abs/1005.4075}{{\ttfamily arXiv:1005.4075 [hep-th]}}.

\bibitem{Banks:2010zn}
T.~Banks and N.~Seiberg, ``{Symmetries and Strings in Field Theory and
  Gravity},'' \href{http://dx.doi.org/10.1103/PhysRevD.83.084019}{{\em
  Phys.Rev.} {\bfseries D83} (2011) 084019},
\href{http://arxiv.org/abs/1011.5120}{{\ttfamily arXiv:1011.5120 [hep-th]}}.

\end{thebibliography}\endgroup
\end{document}